# An X-Ray Burst from a Magnetar Enlightening the Mechanism of Fast Radio Bursts

M. Tavani[1,2], C. Casentini[1,3], A. Ursi[1], F. Verrecchia[4, 5], A. Addis[6], L.A. Antonelli[5], A. Argan[1], G. Barbiellini[7,8], L. Baroncelli[6], G. Bernardi[9,10], G. Bianchi[9], A. Bulgarelli[6], P. Caraveo[11], M. Cardillo[1], P.W. Cattaneo[12], A.W. Chen[13], E. Costa[1], E. Del Monte[1], G. Di Cocco[6], G. Di Persio[1], I. Donnarumma[14], Y. Evangelista[1], M. Feroci[1], A. Ferrari[16,17], V. Fioretti[6], F. Fuschino[6], M. Galli[18], F. Gianotti[6], A. Giuliani[11], C. Labanti[6], F. Lazzarotto[1], P. Lipari[19,20], F. Longo[7,8], F. Lucarelli[4, 5], A. Magro[21], M. Marisaldi[23,6], S. Mereghetti[11], E. Morelli[6], A. Morselli[3], G. Naldi[9], L. Pacciani[1], N. Parmiggiani[6], F. Paoletti[24], A. Pellizzoni[25], M. Perri[4,5], F. Perotti[11], G. Piano[1], P. Picozza[2,3], M. Pilia[25], C. Pittori[4, 5], S. Puccetti[14], G. Pupillo[9], M. Rapisarda[11], A. Rappoldi[12], A. Rubini[1], G. Setti[9,26], P. Soffitta[1], M. Trifoglio[6], A. Trois[25], S. Vercellone[27], V. Vittorini[1], P. Giommi[4,28], F. D'Amico[14]

1. *INAF-IAPS Roma, via del Fosso del Cavaliere 100, I-00133 Roma, Italy*
2. *Università di Roma Tor Vergata, Dip. di Fisica, via della Ricerca Scientifica 1, I-00133 Roma, Italy*
3. *INFN Roma Tor Vergata, via della Ricerca Scientifica 1, I-00133 Roma, Italy*
4. *ASI Space Science Data Center, SSDC-ASI, via del Politecnico snc, I-00133 Roma, Italy*
5. *INAF - Osservatorio Astronomico di Roma, I-00078 Monte Porzio Catone, Italy*
6. *INAF-IASF Bologna, via Gobetti 101, I-40129 Bologna, Italy*
7. *Dip. Fisica, Università di Trieste, via A. Valerio 2, I-34127 Trieste, Italy*
8. *INFN Trieste, Padriciano 99, I-34012 Trieste, Italy*
9. *INAF-IRA, via Piero Gobetti 101, 40129 Bologna, Italy*
10. *Dep. of Physics and Electronics, Rhodes University, PO Box 94, Grahamstown, 6140, South Africa*
11. *INAF-IASF Milano, via E. Bassini 15, I-20133 Milano, Italy*
12. *INFN Pavia, via A. Bassi 6, I-27100 Pavia, Italy*
13. *School of Physics, Wits University, Johannesburg, South Africa*
14. *Agenzia Spaziale Italiana, via del Politecnico snc, I-00133 Roma, Italy*
15. *ENEA Frascati, via Enrico Fermi 45, I-00044 Frascati, Italy*
16. *Dipartimento di Fisica, Università di Torino, I-10133 Torino, Italy*
17. *Consorzio Interuniversitario Fisica Spaziale (CIFS), villa Gualino - v.le Settimio Severo 63, I-10133 Torino, Italy*
18. *ENEA Bologna, via don Fiammelli 2, I-40128 Bologna, Italy*
19. *INFN Roma 1, p. le Aldo Moro 2, I-00185 Roma, Italy*
20. *Dip. Fisica, Università La Sapienza, p. le Aldo Moro 2, I-00185 Roma, Italy*
21. *Institute of Space Sciences and Astronomy (ISSA), University of Malta, Msida MSD 2080, Malta*
22. *CNR, IMIP, Montelibretti (Rome), Italy*
23. *Birkeland Centre for Space Science, Department of Physics and Technology, University of Bergen, Bergen, Norway*
24. *East Windsor RSD, 25A Leshin Lane, Hightstown, NJ 08520, USA*
25. *INAF - Osservatorio Astron. Cagliari, Poggio dei Pini, I-09012 Capoterra, Italy*
26. *Dipartimento di Fisica e Astronomia, Università di Bologna, Via Gobetti 93/2, I-40129 Bologna, Italy*
27. *INAF Osservatorio Astronomico di Brera, v. E. Bianchi 46, I-23807 Merate, Italy*
28. *Institute for Advanced Study, Technische Universitaet Muenchen, Lichtenbergstrasse 2A, D-85748 Garching bei Munchen, Germany*

**Fast radio bursts (FRBs) [1,3] are short (millisecond) radio pulses originating from enigmatic sources at extragalactic distances so far lacking a detection in other energy bands. Magnetized neutron stars (magnetars [4]) have been considered as the sources powering the FRBs, but the connection is controversial because of differing energetics and the lack of radio and X-ray detections with similar characteristics in the two classes. We report here the detection by the AGILE satellite on April 28, 2020 of an X-ray burst in coincidence with the very bright radio burst from the Galactic magnetar SGR 1935+2154 [5,6]. The burst detected by AGILE in the hard X-ray band (18-60 keV) lasts about 0.5 seconds, it is spectrally cutoff above 80 keV, and implies an isotropically emitted energy ~ $10^{40}$ erg. This event is remarkable in many ways: it shows for the first time that a magnetar can produce X-ray bursts in coincidence with FRB-like radio bursts; it also suggests that FRBs associated with magnetars may emit X-ray bursts of both magnetospheric and radio-pulse types that may be discovered in nearby sources. Guided by this detection, we discuss SGR 1935+2154 in the context of FRBs, and especially focus on the class of repeating-FRBs. Based on energetics, magnetars with fields B ~ $10^{15}$ G may power the majority of repeating-FRBs. Nearby repeating-FRBs offer a unique occasion to consolidate the FRB-magnetar connection, and we present new data on the X-ray monitoring of nearby FRBs. Our detection enlightens and constrains the physical process leading to FRBs: contrary to previous expectations, high-brightness temperature radio emission coexists with spectrally-cutoff X-ray radiation.**

Soft gamma-ray repeaters (SGRs) are Galactic compact sources occasionally becoming active in producing tens or hundreds of X-ray/hard X-ray bursts within weeks or months. They are believed to be neutron stars classified as magnetars [4] releasing energy because of instabilities originating from their magnetospheres. SGR 1935+2154 is a magnetar first detected by the *Swift* satellite in 2014 [7] and possibly associated with the SNR G57.2+0.8 at 12.5 kpc [8]. It rotates with a spin period P = 3.24 sec, and it has a dipole magnetic field $B_m = 2 \times 10^{14}$ G as deduced from its spindown properties [9]. The source recently started a period of X-ray bursting on April 22, 2020 [10].

During routine operations in April 2020, the AGILE satellite [11] detected many tens of hard X-ray bursts originating from SGR 1935+2154. The satellite instrument is currently operating with the fully operational MCAL detector sensitive in the 400 keV – 100 MeV, and with Super-AGILE (SA), Anti-coincidence (AC) and gamma-ray imager (GRID) ratemeters (RMs) being sensitive in the energy bands 18-60 keV, 80-200 keV, and 20 keV – 1 MeV, respectively. Figure 1 shows a sample of a sequence of many X-ray bursts attributed to SGR 1935+2154 (a ''burst forest'') detected by AGILE [12] during 7 hours from Apr. 27, 2020 18:31 UT until Apr. 28, 2020 01:45. These X-ray bursts have durations ranging from a fraction of a second to several seconds, and are quite bright, with fluences in the range of $10^{-5} – 10^{-6}$ erg/cm$^2$ (18-60 keV).

Among the several bursts detected on April 28, 2020, one X-ray burst is particularly relevant [13]: it was detected on Apr. 28, 2020 at $T_0$ = 14:34:24.000 +/- 0.256 UT by the Super-A RM in temporal coincidence with the super-bright (double) radio pulse from SGR 1935+2154 revealed at 400-800 MHz by CHIME/FRB [5] and at 1.4 GHz by STARE2 [6]. The timing of the radio event provided by CHIME/FRB (topocentric arrival time at 400 MHz) agrees with that of the AGILE X-ray burst once the time difference of 8.6 s is taken into account as dispersive delay of radio waves along a path with dispersion measure $DM$ = 332.8 pc cm$^{-3}$ [5]. Given the off-axis detection of CHIME/FRB, their radio fluence measurement at 400-800 MHz of a few kJy ms is uncertain. The STARE2 team reports the remarkable lower limit on the fluence at 1.4 GHz of 1.5 MJy ms. Taken at their face values, the CHIME and STARE2 fluences would imply a strong absorption at lower frequencies with important implications on the physical properties of the emitting plasma.

Compared with other X-ray bursts of the SGR 1935+2154 “forest”, the X-ray burst coincident with the intense radio pulse is quite weak in the AGILE data as a consequence of intrinsic faintness of the event in the 18-60 keV band and geometry (event at 121 deg off-axis angle). Figure 2 shows the burst light curve in the 18-60 keV energy band. The event lasts no longer than 0.5 s and has a S/N ratio of 3.8 sigma over the background; its false alarm rate is $3 \times 10^{-3}$ Hz, and the post-trial significance is 2.9 sigma (see Methods). No significant simultaneous detection is obtained from MCAL and AC data, indicating a burst with a relatively soft spectrum and substantially reduced emission above 80 keV. This X-ray burst detected by AGILE is in temporal and spectral agreement with the simultaneous detection by several space instruments (Integral [14], Konus-Wind [15], Insight/HMXT [16]). The fluence in the 18-60 keV band is $5 \times 10^{-7}$ erg cm$^{-2}$ that corresponds to an isotropic equivalent emitted energy of $E_{X,iso} = 8.1 \times 10^{39}$ erg for an assumed SGR 1935+2154 distance of 12 kpc [8]. The event is short and has a spectral cutoff near 50-80 keV (also confirmed in [15]).

When compared with the properties of the radio burst, interesting features emerge. The lower limit on the radio fluence translates into an isotropically radiated energy at 1.4 GHz of $E_{R,iso} \geq 10^{36}$ erg. The ratio $\xi$ between burst X-ray and radio energies is then $\xi = E_{X,iso}/E_{R,iso} \leq 10^4$, a value that is not dissimilar from what observed from impulsive mechanisms of particle acceleration and radiation in astrophysical shocks and unstable systems. What makes this X-ray event remarkable is its association with an FRB-like pulse of coherent radio emission lasting a few milliseconds, a phenomenon never detected before from a magnetar system. Our detection clearly shows that coherent radio emission with very large brightness temperature temporally if not spatially coexists with an energy release of a short burst of X-rays up to 50-80 keV. Locations of the radio and X-ray emissions can be the same or disjoint, leading to interesting possibilities as we discuss below.

Following the radio detection, SGR 1935+2154 was observed by our group at 408 MHz with the Northern Cross Telescope [17] on April 30, May 1 - 8, and May 12 - 16, 2020. The source was observed each day at transit, from 04:00:41 UT till 04:29:09 UT.

Data were acquired with a time resolution of 138 μs and frequency resolution of 12 kHz over a 16-MHz band; they were searched for bursts as large as 700 ms using a Heimdall-based pipeline in a *DM* range $0 < DM < 1000$ pc cm$^{-3}$ (i. e. [18]). No detection was found above 7-sigma upper limits on the fluence of a potential radio burst of ~1.4 Jy ms.

We can consider the energetics of the radio/X-ray burst from SGR 1935+2154 in the context of FRBs. We focus first on the properties of ''repeating-FRBs'' (R-FRBs) that have been recently detected [19,20,21]. Figure 3 shows the distribution of burst energies emitted in the radio band as a function of burst duration for all known R-FRB sources. For comparison, we also add to the plot the $E_{R,iso}$ of SGR 1935+2154. For the first time, a sensible comparison among the different intrinsic energetics between R-FRBs and a magnetar can be done. Remarkably, $E_{R,iso}$ of SGR 1935+2154 is the same as one of the weakest detected R-FRB events, and it does not differ by more than 2 orders of magnitude compared with the bulk of the population. R-FRBs appear to be more energetic than the radio pulse from SGR 1935+2154. However, if the average energy of the R-FRB radio bursts is related with the magnetar magnetic field dissipation, a larger factor of $10^2$ in $E_{R,iso}$ magnitude can be accounted for by values of magnetic fields $B_m$ larger than ~ 10 compared to that of SGR 1935+2154 field, that is $B_m = 10^{15}$ G. Thus the magnetar hypothesis for R-FRBs is plausible from the energetic point of view.

Regarding the X-ray emission, extrapolating from Figure 3 and assuming the same $\xi$ as for SGR 1935+2154 we would expect from the R-FRBs emission of 1-s X-ray bursts related to radio bursts of energies near $10^{42}$ erg. The expected energy flux from this type of X-ray bursts is too low to be detected by current space instruments. However, this is not the only type of X-ray bursts that magnetars can produce. As shown by Galactic SGRs, much more intense X-ray bursts can be emitted (as reviewed in [4]). Restricting our analysis to SGR 1935+2154, there appear to be two types of X-ray bursts emitted by this magnetar. *Type-1 bursts*: associated with intense coherent radio emission, of substantially lower intensity as compared to the most luminous X-ray bursts of the "forest type"; *type-2 bursts*, standard SGR bursts that can release $10^{40}$ - $10^{43}$ erg of energy in 0.1-10 s duration bursts as detected during the "burst forest" episode of Figure 1. These two types of bursts can have different origins, either close to the magnetar surface (e.g., [22,23]) or relatively far from the magnetosphere (e.g., [24,25]). In addition to type-2 bursts (with no associated intense radio pulses) and type-1 (with radio bursts), SGRs are known to emit also rare ''giant'' X-ray/gamma-ray bursts of the type observed from SGR 1806-20 in 2004 that released more than $10^{46}$ erg in 200 ms up to MeV energies [26,27] with no simultaneous radio emission [28]. We call these rare events *type-3 bursts.* Furthermore, active SGRs radiate long-lived X-ray enhanced emission about 100 times more intense than the quiescent state; in case of SGR 1935+2154, the current X-ray luminosity is $8 \times 10^{34}$ erg $s^{-1}$ [29].

If X-ray bursts are emitted by the R-FRBs of Figure 3, interesting consequences follow. Figure 4 shows the current upper limits obtained by AGILE for one of the most interesting R-FRBs, the nearby FRB180916.J0158+65 (hereafter FRB 180916) at the distance of about 150 Mpc [30]; this R-FRB is not only repeating but also periodic with a period of 16.35 days [31]. The purple diamond in Figure 4 shows the expected flux from a type-1 event similar to SGR 1935+2154 if put at the distance of 150 Mpc. Type-2 and type-3 bursts would have larger fluxes from 2 to 6 orders of magnitude, respectively. It is then clear that X-ray monitoring of nearby R-FRBs can reveal bright magnetar-like X-ray bursts, and this task is within the reach of current X-ray instruments. Much harder is the detection of type-1 bursts that requires a source substantially closer than FRB 180916. Figure 4 shows that type-1 bursts similar to SGR 1935+2154 are undetectable unless a very large reservoir of magnetic energy is available for X-ray bursting. Magnetars associable to known R-FRBs most likely have magnetic fields near
$B_m = 10^{15}$ G as indicated above. Magnetars with larger intrinsic magnetic fields $B_m = 10^{16}$ G have also been advocated for FRBs (e.g., [25]). We therefore might expect a broad range of type-1 magnitude and fluxes depending on the source distance, and detection by current X-ray monitoring instruments might still be possible.

The periodic-repeater FRB 180916 is subject to an ongoing monitoring campaign in radio [18], X-ray and gamma-ray energies [32]. Figure 5 shows the results of the X-ray monitoring of the periodic R-FRB 180916 that constrains the average luminosity of long-lived outbursts during active intervals over many cycles to be less than $10^{41}$ erg $s^{-1}$. *Chandra* observations in December 2019 obtained a lower value ($2 \times 10^{40}$ erg $s^{-1}$) [34]. Future improvements of the short timescale limits will be possible by special observation modes. Given the possible binary nature of the radio burst periodicity of R-FRB 180916, these limits are significant to constrain shock interactions within the binary. Another source of interest is FRB 181030 which fares the smallest observed DM among the R-FRBs [21]. Once the Galactic and possible host galaxy contributions are subtracted, the excess DM and resulting distance are even less than those of FRB 180916 [33]. It is then possible to detect X-ray bursts from this source by future monitoring.

It is uncertain whether a single or different types of compact objects are related with the FRB phenomenon (e.g., [3,35]). We made the case of a closeness in energetics and overall phenomenology between SGR 1935+2154 and R-FRBs. Whether this connection applies to the rest of the FRB population is uncertain. Figure 6 in Methods shows the energy of radio bursts of all FRBs (repeating and non-repeating ones) as compared with the radio burst from SGR 1935+2154. The challenge of explaining 6 orders of magnitude of radio emission by models involving only magnetars is severe. More than one type of compact sources might be involved [35]. Beaming and/or plasma lensing [36] are possible features of the emission that can alleviate many if not all of these issues. In order to provide a useful contribution to the energetics problem, the beaming/lensing factors *b and l* should be $b, l \leq 10^{-2}$. All quantities considered above for FRBs should be multiplied by the factors *b,l* to obtain their intrinsic values. If this correction were applied to the R-FRB energies of Figure 3, the SGR 1935+2154 radio burst energy (assumed isotropically emitted) would fit those of nearby FRBs. More data are necessary to resolve this important issue.

The detection of X-rays simultaneous with the very bright radio burst from SGR 1935+2154 is relevant to make progress in understanding the physical mechanism of FRBs. Type-2 and type-3 X-ray bursts of SGRs are believed to be emitted by magnetar events near the magnetosphere involving magnetic field re-arrangements and Alfven wave emission and consequent particle heating (e.g., [37]). The location of type-1 X-ray bursts is uncertain. In case of SGR 1935+2154 the temporal coincidence with an FRB-like radio burst is indicative of a space closeness of the two phenomena.

Modelling of FRB radio bursts focused on coherent plasma radiation processes; a possible mechanism is electron cyclotron maser emission [38] in the context of relativistic shocks of magnetar outflows (e.g., [24,39,40]). Radiation in the GHz range can be emitted with FRB characteristics as a shock precursor of a relativistic fluid interacts with an upstream magnetized environment providing target material for decelerating the outflow. However, the cyclotron maser mechanism is intrinsically narrow band, with $\Delta\omega/\omega$ = a few, and no X-rays are expected by the precursor itself [40]. It is interesting to note that perpendicular shocks of the type believed to apply to pulsars and FRB systems, do not produce power-law tails in purely electron-positron outflows in the absence of ions [38]. Rather, they lead to energy randomization as a consequence of cyclotron emission producing downstream quasi-thermal particle distributions. These types of shocks therefore do not develop a full-blown non-thermal acceleration with power-law distributions to large energies [38]. Therefore, quasi-thermal electron-positron spectra may coexist downstream with a temporary precursor propagating perpendicularly to the local magnetic field. In principle, an upstream-generated coherent radio pulse can coexist with incoherent X-ray emission generated downstream.

Alternately, the radio/X-ray emission site is close to the magnetar magnetosphere (e.g., [23]). In this case, the X-ray emission should resemble that of type-2 bursts, and radio emission can result from curvature radiation within a coherence length compatible with the magnetar geometry [23,41]. However, the coexistence of the multiple radiation processes considered so far, and the simultaneity of X-ray and radio emissions may be challenging for the models so far proposed.

We conclude stating that repeating-FRBs have observed radio bursts and deduced physical parameters not dissimilar from SGR 1935+2154; magnetars with magnetic fields in the range $10^{15}$-$10^{16}$ G can be the underlying sources, and X-ray observations may reveal soon X-ray bursting activity of nearby FRBs. Magnetars can be considered as strong candidates for explaining R-FRBs; whether magnetar-driven processes can explain the bulk of FRBs is an open question that awaits more observations for its resolution.

## Acknowledgements

AGILE is a space mission of the Italian Space Agency developed and operated with the collaboration of INAF and INFN. Research carried out with partial support by the ASI grants: I/028/12/05, ASI 2014-049-R.0. We thank D.D. Frederiks for useful exchanges.

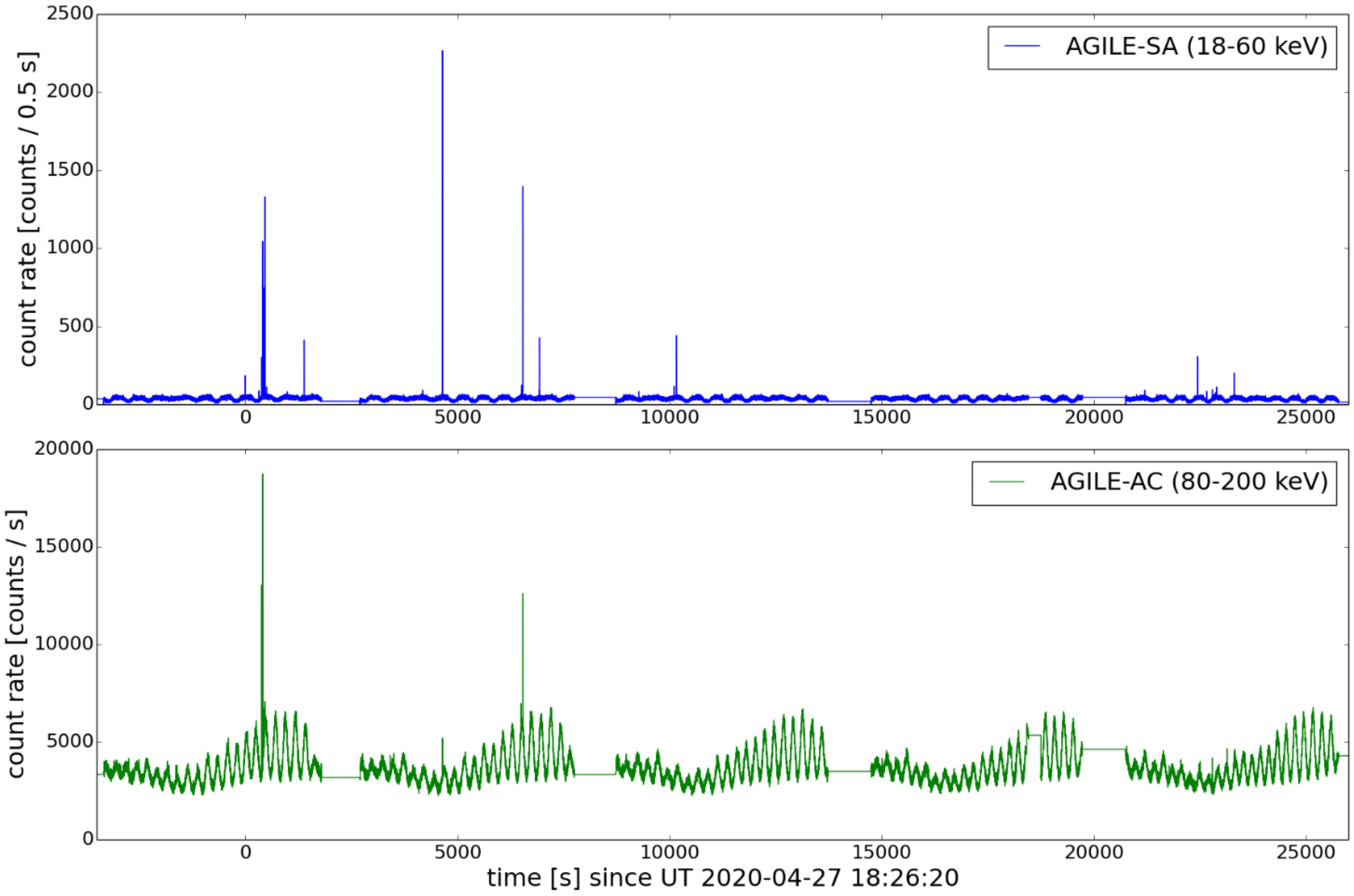

*Figure 1 – AGILE detection of the "forest" of X-ray bursts from SGR 1935+2154. Ratemeter lightcurves of SA data in the 18-60 keV band (top panel) and AC data (80-200 keV) (bottom panel) including the time interval Apr. 27, 2020 18:26 UT and Apr. 28, 2020 01:45 UT. Several X-ray bursts from SGR 1935+2154 are clearly visible. The quasi-sinusoidal variations of the baseline lightcurves are caused by background variations induced by the 7-min spinning of the AGILE satellite.*

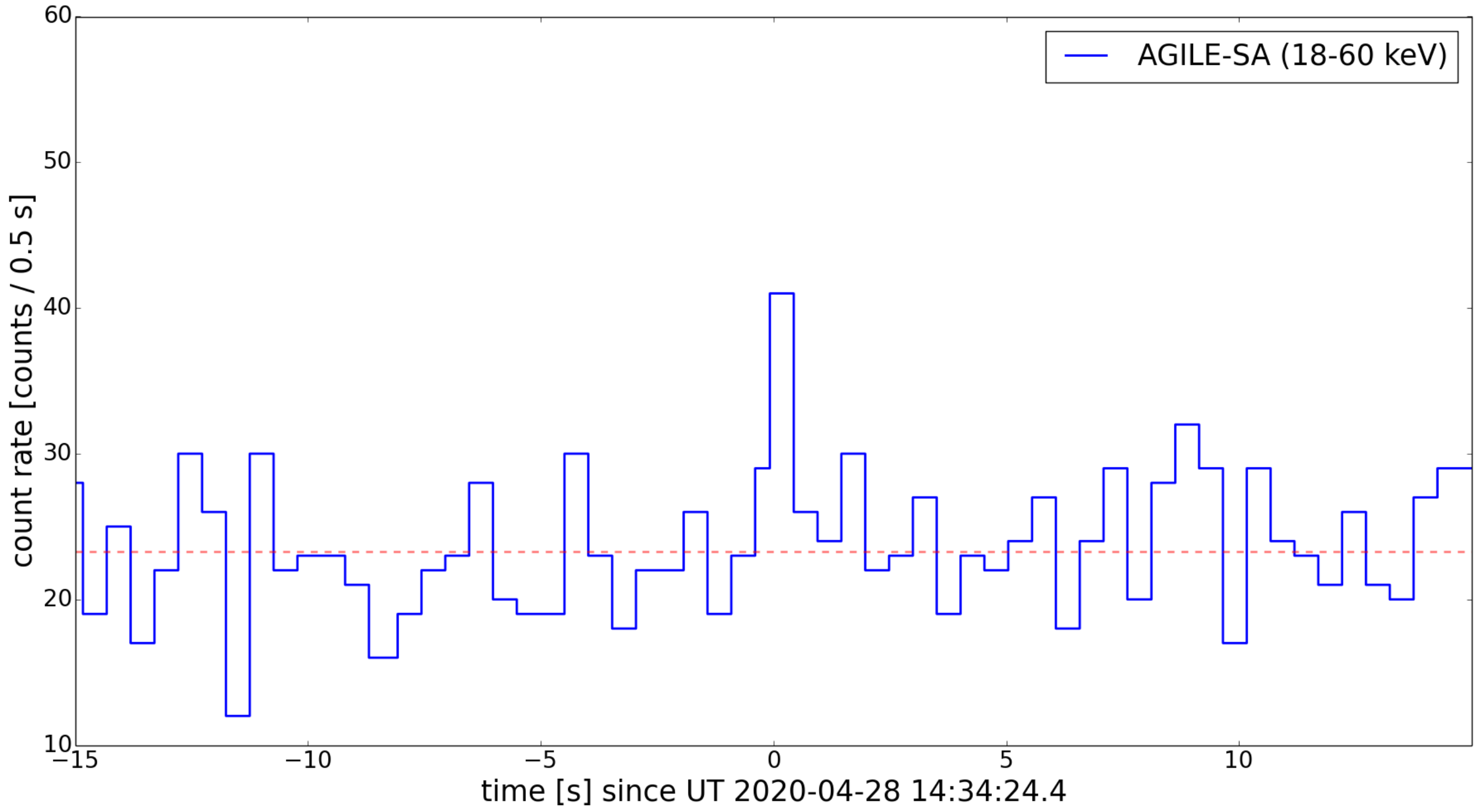

*Figure 2- Detection of the X-ray burst in temporal coincidence with the very intense radio burst from SGR 1935+2154. The panel shows the lightcurve of the AGILE-SA ratemeter with data in the energy range 18-60 keV displayed with 0.5 s binning.*

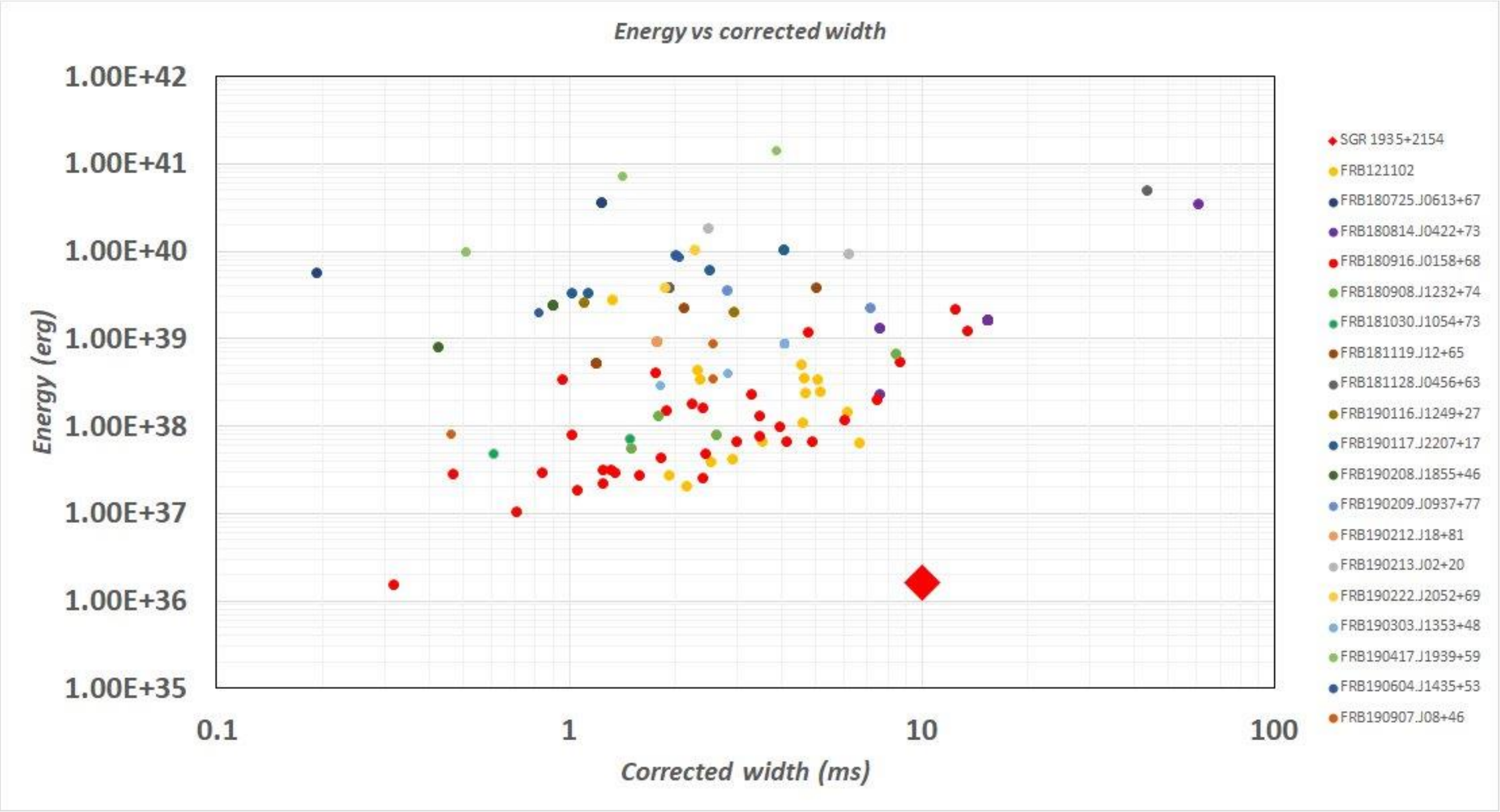

*Figure 3 – Energy of radio bursts detected from the known repeating-FRBs vs. the redshift-corrected time widths of their emission. FRB radio data (typically in the 400 MHz – 1.5 GHz band) and DM information are from ref. [2]. The red diamond marks the observed energy at 1.4 GHz and width of the very intense radio burst from SGR 1935+2154 [6]. The red circles are the measurements of the periodic-repeating FRB 180916 [2,21,30]. The distance of this source is 149 Mpc as from direct measurements [30]. Distances of other repeater-FRBs have been derived from their inter-galactic dispersion measure $DM_{IM}$ once the Galactic disk contribution has been subtracted from the observed DM together with the Galactic halo and host-galaxy contributions (this latter value has been assumed to be equal to $DM_{halo} + DM_{host} = 50 + 50 = 100$ pc cm$^{-3}$, in good agreement with the direct measurement of FRB 180916). See Methods for more details.*

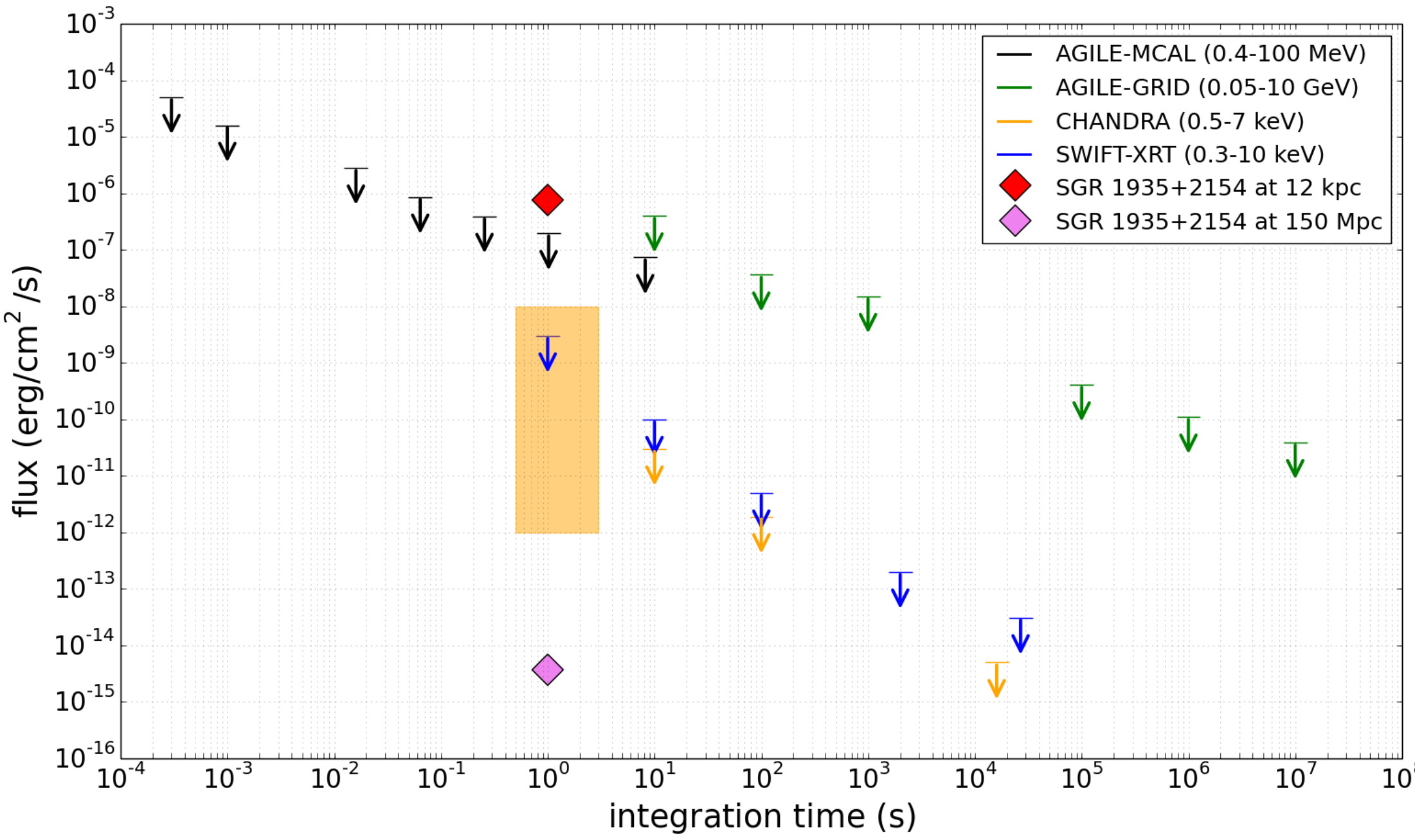

*Figure 4 – High-energy (X-ray and gamma-ray) flux upper limits as a function of integration timescales of observations obtained by AGILE [32,33], Swift [32] and Chandra [34] satellites in monitoring the nearby repeating FRB 180916. Data refer to different detectors in the energy ranges specified in the figure inset. We re-evaluated the upper limits of the Swift and Chandra instruments for integration timescales less than $10^3$ s. We also display the flux of the type-1 X-ray burst from SGR 1935+2154 (red diamond) and for the same source at a 150 Mpc distance (purple diamond). Extrapolations of Swift and Chandra upper limits to 1 s integrations are obtained depending on the so-far adopted observation modes. The shaded region in orange shows the flux range for possible type-1 and type-2 X-ray outbursts from nearby magnetars associated with FRBs. A closer FRB source of a type similar to FRB 180916 and capable of producing X-ray bursts $10^3 – 10^4$ times more luminous than the type-1 burst from SGR 1935+2154 might be detectable by future X-ray monitoring instruments.*

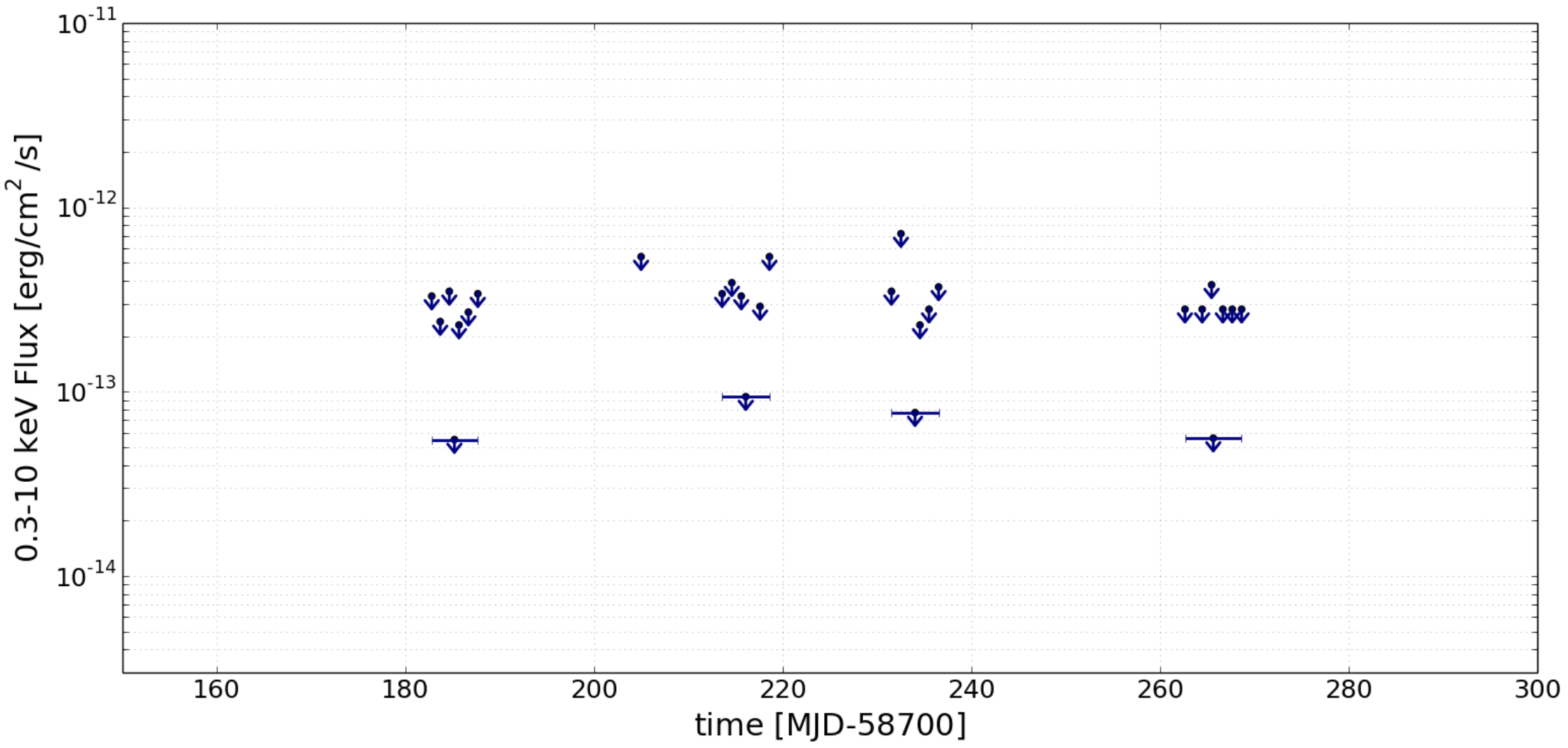

*Figure 5 – X-ray monitoring by Swift-XRT of the periodic R-FRB 180916 (period of 16.3 days) during 5-day active time intervals of expected radio bursting (updated data from [32]). Small arrows indicate flux upper limits in the range 0.3-10 keV obtained for XRT pointings lasting 1-2 ks. The lower value upper limits are the results of the sums of the individual pointings for different active time windows.*

# Methods

## AGILE Scientific Ratemeters (RMs)

Data acquired by all AGILE detectors (i.e., GRID, MCAL, SA, and AC) are continuously collected and recorded in telemetry, used to build broad-band energy spectra in different energy ranges. The GRID, MCAL, and AC RM data are acquired with a 1.024 s time resolution, whereas the SA RM data are acquired with a 0.512 s time resolution.

The RM data provide a continuous monitoring of the hard X- and gamma-ray background, allowing to investigate its modulation in time and space through orbital phases. So far, the AGILE RMs clearly detected a large number of high-energy transients, such as GRBs, SGRs, and solar flares.

The AGILE scientific RMs are routinely calibrated, using GRBs reported by the Inter-Planetary Network (IPN), and comparing the count rate and spectra acquired in the different energy bands with those provided on the same events by other space missions.

## Statistical significance of the X-ray burst on April 28, 2020

The hard X-ray burst detected by the AGILE-SA RM at T0 = 2020-04-28 14:34:24.400 ± 0.245 s (UT) has a Signal-to-Noise Ratio (SNR) of ~3.8 above a background rate of 45 Hz (in the 18-60 keV energy range). We estimated the occurrence of events with a SNR ≥ 3.8 in the AGILE-SA RMs, within ±2.5 ks around the T0, and obtaining a False Alarm Rate, FAR = $3\times10^{-3}$ Hz.

Our signal occurs within ±0.25 s from the X-ray event detected by INTEGRAL/IBAS [14], Konus-Wind [15], and Insight-HXMT [16], compatible with the de-dispersed arrival topocentric time of the CHIME/FRB radio detection [5], at infinite frequency.

It is important to determine the False Alarm Probability (FAP) of a simultaneous occurrence of this event: this value strongly depends on the SNR, the FAR, the time interval between the nominal radio signal and the X-ray burst, the bin-width, and the search window adopted for the search. We use the formula [42],

$$\mathrm{FAP} = \mathrm{FAR} \times \delta t \times \left[1 + \ln\left(\frac{\Delta t}{t_{bin}}\right)\right]$$

Adopting the values, FAR $= 0.003$ Hz, a time distance $\delta t = 0.25$ s, a search window Δt = 30 s, and a bin-width $t_{bin} = 0.5$, we obtain a post-trial probability $P = 0.0039 = 2.9$ σ.

## FRB distances and intrinsic temporal widths

We use data on all FRBs observed to date through the FRBCAT[1] Catalog [2], and the CHIME/FRB experiment online catalogue[2], with modifications for updated information.

We evaluate the FRBs distances starting from the catalogued *DM* values. We considered four different contributors to the estimated *DM*: Galactic (Milky-Way disk, $DM_{(GAL)}$), Galactic halo (Milky-Way halo, $DM_{(HALO)}$), inter-galactic, $DM_{(IG)}$ and host galaxy, $DM_{(HOST)}$. $DM_{(GAL)}$ value was obtained by the catalogues. From literature we find $DM_{(HALO)}$ to be in the range 50-80 pc cm$^{-3}$ [44]. As in our

[1] available at the web page https://www.frbcat.org
[2] available at the web page https://www.chime-frb.ca

galaxy, we have a disk and a halo contribution also for the host galaxy. We sum the effects of these two contributions into a single value. In reasonable agreement with the R-FRB 180916 localization [30], we assume that $DM_{(HALO)} + DM_{(HOST)}$ = 100 pc cm$^{-3}$ for every FRB source with the exception of a few FRBs for which the resulting value of $DM_{(IG)}$ is negative. In these few cases, we adopt different values for $DM_{(HALO)} + DM_{(HOST)}$. We then estimate $DM_{(IG)}$ by subtracting Galactic and host contributions from the total value: $DM_{(IG)} = DM - DM_{(GAL)} - DM_{(HALO)} - DM_{(HOST)}$. Once $DM_{(IG)}$ is obtained, the FRB source distance is calculated using the relation $DM_{IG} = 900\ z$ pc cm$^{-3}$ [45]. The resulting redshift value is then used to estimate the distance following the cosmology as in [45]. The same value of the redshift is also obtained for determining the intrinsic temporal widths of the FRB radio bursts, $\Delta t_{true} = \Delta t_{obs} /(1+z)$.

It is important to test this procedure for a few critical FRBs of intrinsically small $DM$. A test case is R-FRB 180916 that has a total $DM$ of about 350 pc cm$^{-3}$ [21,1]. This value is expected to be the sum of $DM_{(GAL)}$, $DM_{(HALO)}$, $DM_{(HOST)}$, and $DM_{(IG)}$, which includes both the inter-galactic contribution and that of the host galaxy. $DM_{(GAL)}$ is determined, following [46], to be near 200 pc cm$^{-3}$ [2]. We assume a value of 50 pc cm$^{-3}$ for the halo component, $DM_{(HALO)}$. By subtracting $DM_{(GAL)}$ and $DM_{(HALO)}$ from the total $DM$, we are left with $DM_{excess}$ ~ 100 pc cm$^{-3}$, to be divided between $DM_{(IG)}$ and $DM_{(HOST)}$. Knowing the distance of R-FRB 180916 of 150 Mpc [3], we determine the redshift to be z ~ 0.0337. For such a value of z, following the relation $DM_{(IG)}$-$z$, reported in [45], we obtain $DM_{(IG)}$ ~ 31 pc cm$^{-3}$. So, after the subtraction of $DM_{(IG)}$ from the excess 100 pc cm$^{-3}$, we are left with a value of $DM_{(HOST)}$ of the host galaxy around 69 pc cm$^{-3}$. We therefore determine that, despite uncertainties on the single components, the overall adopted value for the sum of $DM_{(HALO)}$ and $DM_{(HOST)}$ totaling ~ 100 pc cm$^{-3}$ may be reasonable, and the uncertainties on the $DM_{(IG)}$ may be of the order of several tens of pc cm$^{-3}$.

Let us consider a case for which this procedure fails. The values for the R-FRB 181030 are even smaller than in the previous case; if we adopted $DM_{(HALO)} + DM_{(HOST)}$ = 100 pc cm$^{-3}$ we would obtain a negative value of $DM_{(IG)}$ from a total value $DM$ = 103.5 pc cm$^{-3}$ and a Galactic plane contribution estimated to be $DM_{(GAL)}$ = 30 pc cm$^{-3}$ [21]. In this case, we adopt $DM_{(HALO)} + DM_{(HOST)}$ = 60 pc cm$^{-3}$ which leads to $DM_{(IG)}$ = 13.5 pc cm$^{-3}$, corresponding to z ~ 0.015 (around 65 Mpc). Clearly this value is uncertain and can be considered as an upper limit to the true distance of R-FRB 181030.

In order to evaluate the uncertainties in the determination of the $DM_{(IG)}$, we have checked the redshift values obtained by our method with the known redshifts determined for five FRBs. Table 1 summarizes the assumptions and the resulting redshifts obtained here. Table 2 shows the assumptions and the measured redshifts of the five FRBs as presented in the literature [30,47, 48,49,50]. Our method based on the assumption $DM_{(HALO)} + DM_{(HOST)}$ = 100 pc cm$^{-3}$ provides values closer to the measured ones within a few percent for FRB181112 and FRB190523. Specific Milky Way halo plus host galaxy contributions may be different from our general assumption. A noticeable case is FRB180924 for which the measured host galaxy redshift of z = 0.321 [48] corresponds to a $DM_{(IG)}$ = 288 pc cm$^{-3}$ against our deduced value of 221 pc cm$^{-3}$. The corresponding redshifts differ by 30% as a consequence of a value of $DM_{(HALO)} + DM_{(HOST)}$ anomalously smaller than 100 pc cm$^{-3}$. The opposite case is provided by FRB121102.

We conclude that, despite the uncertainties on the individual contributions for $DM_{(HALO)}$ and $DM_{(HOST)}$ and source-specific peculiarities, the overall energy determinations of Figures 3 and 6 are reasonable within factors of order 2-4.

**Table 1**

| Source | Ref. | DM (pc cm$^{-3}$) | $DM_{(GAL)}$ (pc cm$^{-3}$) | $DM_{(HALO)}$ (pc cm$^{-3}$) | $DM_{(HOST)}$ (pc cm$^{-3}$) | $DM_{(IG)}$ (pc cm$^{-3}$) | $z_{est}$ |
|---|---|---|---|---|---|---|---|
| FRB121102 | this paper | 558.1 ± 3.3 | 188 | 50 | 50 | 270 | 0.300 |
| FRB180916 | this paper | 348.8 ± 0.1 | 199 | 50 | 50 | 50 | 0.056 (*) |
| FRB180924 | this paper | 361.42 ± 0.06 | 40 | 50 | 50 | 221 | 0.246 |
| FRB181112 | this paper | 589.27 ± 0.03 | 42 | 50 | 50 | 447 | 0.490 |
| FRB190523 | this paper | 760.8 ± 0.6 | 37 | 50 | 50 | 624 | 0.690 |

(*) In Fig. 3 and 6 we used for FRB 180916 the observed redshift $z_{obs}$ = 0.034 corresponding to a distance of 149 Mpc.

**Table 2**

| Source | Ref. | DM (pc cm$^{-3}$) | $DM_{(GAL)}$ (pc cm$^{-3}$) | $DM_{(HALO)}$ (pc cm$^{-3}$) | $DM_{(HOST)}$ (pc cm$^{-3}$) | $DM_{(IG)}$ (*) (pc cm$^{-3}$) | $z_{obs}$ |
|---|---|---|---|---|---|---|---|
| FRB121102 | 47 | 558.1 ± 3.3 | 188 | 50 | 150 | 171 | 0.19 |
| FRB180916 | 30 | 348.8 ± 0.1 | 199 | 50 – 70 | 48 – 68 | 31 | 0.034 |
| FRB180924 | 48 | 361.42 ± 0.06 | 40 | < 60 | < 30 | 289 | 0.321 |
| FRB181112 | 49 | 589.27 ± 0.03 | 42 | 60 | 55 | 432 | 0.48 |
| FRB190523 | 50 | 760.8 ± 0.6 | 37 | 50 – 80 | 50 – 80 | 594 | 0.66 |

(*) The values of $DM_{(IG)}$ correspond to the observed redshifts $z_{obs}$ given in the next column. The values of $DM_{(HALO)}$ and $DM_{(HOST)}$ are discussed in the reference papers.

**FRB energy estimates**

Once the values of the redshift z are known for the individual FRBs, we proceed to the estimates of the energies of the radio bursts under the assumption of isotropic emission. Alternately, we can evaluate the radio burst energies from the relation $E = F \times DM_{(IG)}^2 \times K \times \Delta t \times \Delta \nu$ with $F$ is the spectral flux in Jy, $\Delta t$ is the intrinsic temporal width in units of 1 ms, the bandwidth is $\Delta\nu$ = 1 GHz, and $DM_{(IG)}$ = 34 pc cm$^{-3}$, that is the value of R-FRB at the known distance of 150 Mpc. The factor $K = 2 \times 10^{34}$ erg pc$^{-1}$ cm$^{3}$ Jy$^{-1}$ ms$^{-1}$ GHz$^{-1}$ is the conversion factor from fluence to energy. In practice, we use this value of $K$ to convert all the FRB fluences to energies obtaining the plots showed in Figures 3 and 6. From the previous analysis, energy determinations are uncertain within factors of 2-4.

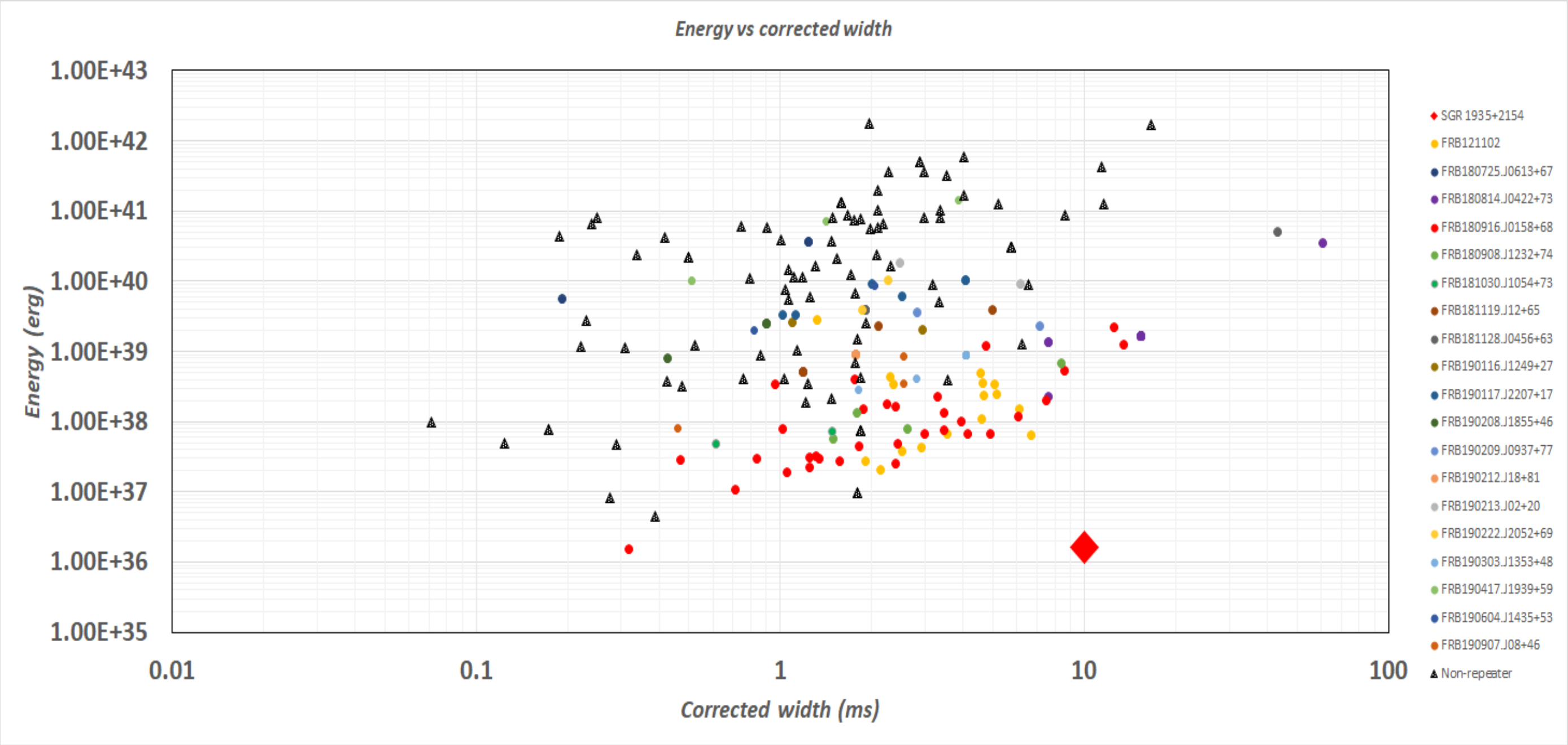

*Figure 6 – Energy of the FRB radio bursts vs. their redshift-corrected intrinsic time widths. See text for details on the method for obtaining distances from measured DMs and energies. The value of the energy emitted in the radio band by SGR 1935+2154 is marked by a red diamond. In determining the energies, we use the appropriate frequency bandwidths of different searches.*

## Additional References